\documentstyle[11pt,aaspp4]{article}
\begin{document}
\newcommand{\nan}{Nan\c{c}ay}
\newcommand{\skp}{\mbox{  }\\}
\newcommand{\kms}{\mbox{km s$^{-1}$}}
\newcommand{\etal}{{\it et al.}}
\newcommand{\msun}{\mbox{${\cal M}_\odot$}}
\newcommand{\bmv}{\mbox{B--V}}
\newcommand{\bmr}{\mbox{B--R}}
\newcommand{\mhi}{\mbox{${\cal M}_{\HIss}$}}
\newcommand{\HI}{\mbox{H\,{\sc i}}}
\newcommand{\HIbf}{\mbox{H\hspace{0.155 em}{\scriptsize \bf I}}}
\newcommand{\HIit}{\mbox{H\hspace{0.155 em}{\scriptsize \it I}}}
\newcommand{\HIss}{{\rm H\hspace{0.04 em}\scriptscriptstyle I}}
\newcommand{\HII}{\mbox{H\,{\sc ii}}}
\newcommand{\dark}{$\frac{{\cal M}_{HI}}{L_{V}}$}
\newcommand{\am}[2]{$#1'\,\hspace{-1.7mm}.\hspace{.0mm}#2$}
\tighten


\title{An Exploration of the Tully-Fisher Relation for Extreme 
Late-Type Spiral Galaxies}

\author{L. D. Matthews}
\affil{Department of Physics \& Astronomy,
State University of New York at Stony Brook, Stony Brook, NY
11794-3800}
\affil{Electronic mail: matthews@gremlin.ess.sunysb.edu}
\authoraddr{Department of Physics \& Astronomy,
State University of New York at Stony Brook, Stony Brook, New York
11794-3800}

\author{W. van Driel}
\affil{Unit\'e Scientifique  Nan\c{c}ay, CNRS USR B704, Observatoire de 
Paris,
92195 Meudon, France}
\affil{Electronic mail: vandriel@mesioq.obspm.fr}
\authoraddr{Unit\'e Scientifique Nan\c{c}ay, Observatoire de Paris, 
92195
Meudon, France}

\author{J. S. Gallagher, III}
\affil{Department of Astronomy, University of Wisconsin---Madison,
Madison, Wisconsin 53706}
\affil{Electronic mail: jsg@tiger.astro.wisc.edu}

\

\slugcomment{To appear in the November  Astronomical Journal}

\begin{abstract}
In an earlier paper (Matthews \etal\ 1998)
we presented new high-precision \HI\ velocity
width measurements for a sample of 30 extreme late-type spiral
galaxies.  Here
we explore the adherence of those
galaxies, as well as 17 additional extreme late-type spirals, to
the $B$- and $V$-band Tully-Fisher relations defined by a sample of 
local calibrators. In 
both bands we find the mean luminosity at a given 
linewidth for extreme late-type
spirals to lie below that predicted by  standard
Tully-Fisher relations. While many of the extreme late-type spirals do
follow the Tully-Fisher relation to within our observational 
uncertainties, most of these galaxies lie below the normal, linear 
Tully-Fisher relation, and some are underluminous by more than 
2$\sigma$ (i.e. $>$1.16 magnitudes in $V$).
This suggests a possible downward curvature of the Tully-Fisher 
relation for 
some of 
the smallest and faintest  rotationally supported disk galaxies. This
may be a consequence of the increasing prevalence of dark matter in
these systems.
We find the deviation from the Tully-Fisher relation to increase with 
decreasing luminosity and
decreasing optical linear size in our sample, 
implying that the physically
smallest and faintest spirals may be a structurally and kinematically distinct
class of objects.
\end{abstract}

\keywords{galaxies: spiral---galaxies: statistics---dark matter}

\section{Introduction}
The Tully-Fisher (TF) relation (Tully \& Fisher 1977) 
is an empirical correlation
which predicts that the luminosity $L$ of a disk galaxy is
proportional to its maximum rotational velocity  
$V_{max}^\alpha$, where $\alpha$ has been
observationally established to be $\sim$3-4 (e.g., Aaronson \etal\ 1979).
In spite of the frequent use of the TF relation as a distance
indicator, the physical origin of this
relationship is still relatively poorly understood, and 
 it remains unclear whether all rotationally
supported disk galaxies, including late-type spirals and irregulars,
obey a single luminosity-linewidth correlation.  Since the mean physical
parameters
which characterize disk galaxies, including total luminosity, gas
fraction, and bulge size, change along the Hubble
sequence, testing whether or not TF
holds in different classes of disk galaxies can indirectly offer important
information about galactic disk structure and evolution
(e.g., Romanishin \etal\ 1982; Persic \& Salucci 1991; Karachentsev 1991;
Zwaan \etal\ 1995; Rhee 1996; Meurer \etal\ 1996; Matthews \etal\ 1997; 
Bershady 1997) and can help to test recent semi-analytic models 
for disk galaxy formation (e.g., White
\& Frenk 1991, Heyl \etal\ 1995).

In an earlier paper, Matthews \etal\ (1998; hereafter Paper~I)
presented new precise \HI\ linewidth measurements
for a sample of 30 southern,
extreme
late-type spiral galaxies\footnote{Matthews \&
Gallagher (1997) define {\it extreme late-type spirals}
as ``the lowest luminosity, late-type disk galaxies which still exhibit
regular optical structures and centralized light concentrations.''}
 with $V_{h}\le$3000~\kms. These 
objects are Sc-Sm spirals with 
moderate-to-low optical surface brightnesses and which fall at the 
low end for spiral galaxies in terms of properties such as \HI\ 
mass, optical luminosity, and optical size.  $B$- and $V$-band CCD imaging
and photometry of these galaxies was also recently published
by Matthews \& Gallagher (1997; hereafter MG). Together
these new datasets allow us to investigate the TF relation
for extreme late-type spiral galaxies. Findings reported by MG, and in Paper~I,
suggest that some of the smallest spiral galaxies (i.e., 
the extreme late types) appear to be structurally
and kinematically distinct from giant spirals (see also Matthews \& 
Gallagher 1998). Therefore
it is of interest to assess whether or not they follow the same TF
relation.

\section{The Luminosity-Linewidth Relation for Extreme Late-Type
Spiral Galaxies}
Matthews \etal\ (1997) presented a preliminary analysis of the TF
relation for 54 extreme late-type (ELT) spiral galaxies.  
Based on a comparison with a sample 
of high  surface brightness galaxies from 
the Nearby Galaxies Catalog (Tully 1988; hereafter the TNGC), their 
finding was that there may be a curvature 
or an offset in the TF relation among some of the smallest and 
faintest spirals. Here we re-examine this trend more rigorously using the
new, improved linewidths
for the sample of galaxies presented in Paper~I. We consider only those 
27 galaxies with $i>30^{\circ}$. To increase our sample size,
we also include 17 extreme late-type spirals with $i>$35$^{\circ}$ from 
the MG photometric
sample, for which \HI\ linewidths are available in the
literature (Fouqu\'e \etal\ 1990; Gallagher \etal\ 1995). Hereafter we 
collectively refer to these 44 galaxies as the ``combined 
ELT sample''.

\subsection{The ``Combined ELT Sample'' and Corrections
to Observed Quantities}

\subsubsection{Internal and External Extinction Corrections}

For the present analysis, we adopt the apparent $B$ and $V$ magnitudes
of MG for the combined ELT sample galaxies. For consistency with other 
workers, we apply a correction to these magnitudes for Galactic 
extinction, $A_{B}$ (taken from Lauberts \& Valentijn 1989), and
for internal extinction: $A^{i-0}_{B}$=$A^{i}_{B}$ - $A^{0}_{B}$, where
\begin{displaymath}
A^{i}_{B} = -2.5log\left[f\left(1+e^{-\tau sec~i}\right) + 
\left(1 - 2f\right)\left(\frac{1 - e^{-\tau sec~i}}{\tau sec~i}\right)\right],
\end{displaymath}
\noindent $A^{0}_{B}$=0.27, $\tau$=0.55,  $f$=0.25, and $i$ is the galaxy
inclination (Tully \& Fouqu\'e 1985; 
hereafter, TFq). For $i>80^{\circ}$, $A^{i-0}_{B}$=0.67 is adopted
(TFq), and we assume $A_{V}$=$A_{B}$/1.33 (Allen 1963). We note that this 
approximate correction
for $V$ formally applies only the limit of a foreground dust screen (the
``normal extinction'' case; see Calzetti \etal\
1994). However, given the low optical depths considered here and the
uncertain dust geometries in extreme late-type spirals, this 
approximation should be satisfactory (see also Han 1992).
Nonetheless, we caution
that our derived extinction
corrections may be overestimates of the internal extinction for many 
moderate-to-low surface brightness galaxies (Rhee 1996; Pierce 1998; 
Sect.~2.4.2).  If extinction is overestimated, then our derived
luminosities will be too high.

\subsubsection{Distance Determinations}

We have computed absolute magnitudes for the combined
ELT sample galaxies using distances derived via two methods:
from a simple linear Hubble law and 
from galaxy group assignments, using the prescription of Tully \etal\ 
(1992) 
(see Sect.~2.4.5).
To derive the linear Hubble flow distances, we used radial
recessional velocities corrected for motion with respect to the Local Group
according to Sandage \& Tammann (1981):
$V_{cor}=V_{h} + \Delta V$, where $\Delta V = -79\cos(l_{II})\cos(b_{II}) 
+  296\sin(l_{II})cos(b_{II}) - 36\sin(b_{II})$~\kms. Here $V_{h}$ is 
the measured heliocentric 
recessional velocity, and $l_{II}$ and $b_{II}$ are
the Galactic latitude and longitude, respectively. For consistency with 
Pierce \& Tully (1988), we adopt a Hubble 
constant of 85~\kms~Mpc$^{-1}$.

\subsubsection{Corrections to Observed Linewidths}

It is generally argued that in order to obtain a measure of the 
maximum rotational velocity of a disk from its measured global \HI\ profile 
width, $W_{20}$,
some correction to the observed line width should be made
for the effects of broadening due to turbulent (i.e. non-rotational) motions
(e.g., Roberts 1978; Bottinelli \etal\ 1983; 
TFq). A commonly used formulation of this correction,
suggested by TFq, assumes that a linear summation
of rotational and random motions adequately describes the observed
profile widths of giant galaxies, while for dwarfs (i.e., slowly
rotating galaxies with Gaussian-like line profiles), a sum 
in quadrature of the the random and rotational  terms is appropriate. However,
the TFq method has serious shortcomings (Rhee 1996), and
it is not necessarily appropriate for the types of objects
considered here. 

Our combined ELT sample consists of galaxies that appear
{\it disk-like} and  {\it regular}  in optical images (MG). And while many
of the  systems have
fairly narrow \HI\ line profiles, 
all exhibit significant rotational broadening and most are double-peaked 
(Paper~I).
Moreover, nearly ``Gaussian'' profiles can still occur for small, 
non-face-on galaxies supported predominantly by rotation
(e.g., Skillman 1996).
Thus linewidth corrections optimized for ``dwarfs'' are unlikely 
to be appropriate for our sample galaxies. For this reason, we 
adopt the linear
turbulence correction formula suggested by Bottinelli \etal\ (1983):
\begin{displaymath}
W_{20,i,c}=[W_{20,obs} - W_{t,20}]\times({\rm sin}~i)^{-1}, 
\end{displaymath}
\noindent where
$W_{20,obs}$ is the observed profile with at 20\% peak maximum, corrected
for instrumental broadening, $W_{t,20}$ is the  correction term for
turbulence, and the factor of $sin~i$ corrects for disk inclination.
For the combined ELT sample, we use the optically derived inclinations 
from MG.

A second matter of debate for computing linewidth corrections
is the optimum value of $W_{t,20}$.
TFq assume a typical velocity dispersion in the gas of a galaxy 
is $\sigma_{z}$=10~\kms, leading to a derived 
turbulent motion term $W_{t,20}$=38~\kms\ (TFq). However, 
$\sigma_{z}$=10~\kms\ is at the high end for observed \HI\ velocity
dispersions  in the outer disks of either dwarf or spiral galaxies, 
where typically
$\sigma_{z}\sim$~5-10~\kms\ (e.g., Dickey \etal\ 1990; 
Kamphuis \& Sancisi 1993; Rownd \etal\ 1994; 
Skillman 1996; Young \& Lo 1996; C\^ot\'e \etal\ 1997;
Pisano \etal\ 1998). 

From a comparison between global \HI\
linewidths and maximum rotational velocities derived from rotation
curves, Rhee (1996)  statistically derived a smaller value for
the turbulent motion correction term of $W_{t,20}$=20~\kms\
 (corresponding to $\sigma_{z}\sim$5.5~\kms\
in the case of a nearly Gaussian velocity dispersion). Although 
the uncertainty in this term for any individual galaxy can be 
large, Rhee (1996) has shown that $W_{t,20}$=20~\kms\ should be the 
optimum value for recovering
$V_{max}$ for a statistical sample of galaxies; we therefore adopt 
this number in the present work.

\subsection{The Comparison Sample}

As a TF comparison sample, we have chosen the standard
local calibrator sample
of Pierce \& Tully (1992; hereafter PT). We used the absolute
$B$ magnitudes
quoted in their paper (corrected for internal and external extinction
as outlined above), and have derived absolute
$V$ magnitudes using the internal and external extinction-corrected
$B - V$ colors from the RC3 (de Vaucouleurs \etal\ 1991). We derived
$W_{20,i,c}$ values for the PT sample from the raw $W_{20}$ values 
and inclinations quoted by PT.  All corrections we apply to the PT 
sample are necessary for consistency with the corrections applied to 
the combined ELT sample.

Cepheid distances are available for only 6 galaxies in the PT
sample, and PT caution against calibrating the TF 
relation using their full sample. However, we find the 
relation we derive from the full sample of PT galaxies to be statistically 
indistinguishable from that derived from the Cepheid galaxies alone, apart
from its larger scatter.
Therefore for the purpose of our analysis, we consider the full PT sample, 
as it yields a better estimate of the intrinsic scatter one might 
expect in a sample of galaxies for which independent distance 
determinations are not available.

\subsection{The Luminosity-Linewidth Distribution for the Combined
ELT Sample}

Fig.~1~$(a)$ and $(b)$ show plots of the $B$- and $V$-band absolute 
magnitude versus the logarithm of the corrected \HI\ linewidth  for 
the 27 extreme late-type spirals from Paper~I (triangles), the 17
extreme late-type spirals from MG (squares), and the PT comparison
sample (asterisks). The absolute magnitudes for our combined ELT sample 
galaxies are based on distances derived from a  simple 
linear Hubble flow
model (Sect.~2.1.2); other distance determination methods are
considered in Sect.~2.4.5. 

We performed a linear fit to the PT points
using the ordinary least squares bisector method, which minimizes the 
residuals in both $x$ and $y$. For small data sets, this is 
the preferred method for finding the intrinsic relationship between 
two variables
in the case where errors for the
individual data points are not explicitly known (Babu \& Feigelson 1996). 
The results of the fits are: 
$M_{B,i,b}=-[(6.12\pm0.37)\times(log(W_{20,i,c}))] - (3.82\pm0.88)$
 and 
$M_{V,i,b}=-[(6.53\pm0.25)\times(log(W_{20,i,c}))] - (3.56\pm0.58$). 
These 
fits are plotted as solid lines on Fig.~1; the dotted lines
indicate the 2$\sigma$ scatter. Our resulting $B$-band fit differs very
slightly from that of PT, as we used a different method for determining
turbulent motion corrections. PT did
not present a $V$-band TF relation for their data, but for normal
galaxy colors, our $V$-band fit is consistent with the expected offsets 
in both slope and  y-intercept from the PT $B$-band and $R$-band 
calibrations.

Inspection of Fig.~1 reveals that while many of the galaxies from 
the combined ELT sample scatter about the TF relations defined by the PT
galaxies, there is a mean offset of the ELTs
toward fainter magnitudes in both
$B$ and $V$. In fact, a number of the combined ELT 
sample galaxies scatter
more than 2$\sigma$ below the fits to the PT galaxies.
Because our combined ELT sample covers a relatively narrow range
in log$(W_{20,i,c})$,  a linear least squares fit to these data does
not have a well-constrained slope. For this reason, we do not attempt 
to analytically fit our combined
ELT sample, but  instead consider the distribution of the combined ELT 
sample points relative to the TF relation defined by the local calibrator 
samples. 

\subsection{Does the Combined ELT Sample  Obey
the TF Relation to within Observational Uncertainties?}

The mean offsets of the combined ELT sample galaxies
from the standard $B$- and $V$-band TF 
relations are 0.64 and 0.82 magnitudes, 
respectively. The reduced chi-squared values resulting from forcing
the fits to the PT local calibrators to the combined
ELT sample are
$\chi^{2}_{\nu,B}$=1.40 (for $V$) and $\chi^{2}_{\nu,V}$=1.72 (for $B$). 
If we assume the local calibrator fits should adequately 
characterize the combined ELT sample,  
there is only  a 5\% probability of obtaining $\chi^{2}_{\nu,B}$ 
this large or larger from a randomly selected set of points, 
and a 0.1\% 
probability of obtaining or exceeding this value of $\chi^{2}_{\nu,V}$. 
Thus it is highly probable
that the standard TF relation does not represent the true underlying 
distribution for the combined ELT sample data points 
in Fig~1. 
Before we can establish whether this observed offset from TF is real, we 
must rule out observational errors and biases which may account for 
part or all of this effect. We discuss various possibilities in turn.

\subsubsection{Photometric Uncertainties} MG present a comparison of their
measured $B$-band isophotal magnitudes with those in the
ESO Catalogue (Lauberts \& Valentijn 1989) and found a mean offset of
$\sim+$0.2 magnitudes, consistent with offsets from the
ESO magnitude system  found by
other workers (e.g., R\"onnback \& Bergvall 1994; Vader \& Chaboyer 
1994; Vennik \etal\ 1996). Uncertainties in individual magnitude measurements
are typically $<$0.03 magnitudes. 
Therefore photometric calibration
uncertainties are unlikely to contribute a
significant offset from TF for the combined ELT sample.

Still, one concern is that for the faintest 
galaxies in the combined ELT sample, the measured magnitudes may not
adequately represent the true total magnitudes,  since for 
galaxies of very low central surface brightness, a significant 
percentage of the total flux lies in the faint, outer portions of the 
disk and may be unmeasurable on  CCD images 
(e.g., McGaugh \& Bothun 1994). We estimate that such 
corrections would be less than $\sim$0.1 
magnitude for our galaxies (see de Blok \etal\ 1995; Tully \etal\ 
1996), yielding a total uncertainty in our absolute magnitudes of $\sim$0.33 
magnitudes.

\subsubsection{The Use of B and V Photometry and the 
Effects of Internal Extinction} 
In theory, it is preferable
to perform TF analyses using near-infrared (NIR) rather than optical
magnitudes in order to
minimize  magnitude uncertainties due to 
internal and external extinction and to reduce sensitivity to
the effects of recent star formation.  However, for faint galaxies,
the high and variable
sky background in the NIR and the large aperture corrections
which are necessary at these wavelengths place serious limits
on achievable photometric accuracy.  The small
field sizes of most currently available
NIR detectors  can also lead to large flatfielding errors (generally
the dominant source of error in the photometry of faint, large
angular size objects). Increasing sky and 
field star brightnesses as well as flatfielding difficulties at $R$ and $I$
also create problems with using these wavebands
for TF studies of faint galaxies. For these reasons,  $B$-band and 
$V$-band
make reasonable choices for exploring the TF relation for extreme 
late-type
spirals, although investigations at other wavelengths will 
naturally be desirable as 
photometry becomes available.

The need to move to 
longer wavelengths for TF  investigations
is also lessened for extreme late-types spirals because
these galaxies  appear to have fairly low
internal extinction (e.g., McGaugh 1994; 
R\"onnback \& Bergvall 1995; Bergvall \& 
R\"onnback 1995; de~Blok \etal\ 1995; Tully \etal\ 1998). 
One way of seeing that internal extinction is likely to be 
low for many of our combined ELT sample objects
 is to compare their extinction-corrected
$B - V$ colors with those of higher surface brightness galaxies.
Fig.~2 shows a plot of $B - V$ color, derived from our internal and
external extinction-corrected magnitudes, versus log$(W_{20,i,c})$
for our combined ELT sample (squares and triangles). As a comparison,
we show the PT local calibrator galaxies (asterisks) and
a sample of galaxies from the TNGC (crosses). The TNGC sample 
consists of all galaxies
in the TNGC that lie within the 
same redshift range ($V_{h}<$3000~\kms) and same angular 
volume ($3^{h}\le\alpha\le14.5^{h}$, $-18^{\circ}\le\delta\le -38^{\circ}$)
as our combined ELT sample. We see that many extreme late-type
spirals appear to be among the bluest galaxies at a given linewidth, 
although they span a range in color. Since bluer galaxies generally
have less dust and 
younger stellar population mixes, both of these effects  
might be expected to 
cause galaxies to lie {\it above} the standard
TF relationship---i.e. to be too bright for their rotation
speeds. However, Fig.~3 shows that there is no apparent correlation between 
extinction-corrected $B - V$ color and deviation from the TF relation, 
$(\Delta TF)_{V}$, in 
our sample.

Further  inspection shows that the extinction corrections applied
to the combined ELT sample have in some cases
lead to unreasonably blue $B - V$ colors [$(B - V)_{i,b}<$0.1],
implying that the internal extinctions estimated using the method
of TFq  are  {\it overestimates}  for at least some 
extreme late-type spirals. Thus the
TFq extinction corrections appear to have artificially ``brightened''
some of our combined ELT sample galaxies, and worked to
decrease any possible intrinsic offsets from TF. The inappropriateness
of the TFq corrections for very late-type galaxies has also been
cited by Rhee (1996) and Pierce (1998). In addition, these workers note
that the TFq corrections may be {\it underestimates} for some very
bright galaxies, hence further masking possible intrinsic TF offsets
between different galaxy samples or possible curvature in the TF relation.

\subsubsection{Uncertainties in W$_{20}$ Measurements} 

For the 27  galaxies from Paper~I, 
uncertainties in $W_{20}$
range from 1-23~\kms. Similar errors are quoted for the $W_{20}$
values from the literature for the remaining 17 galaxies in the combined
ELT sample, as well as for the PT calibrators. This should lead 
to maximum errors in predicted TF magnitudes
of  no more than a few tenths of a magnitude for individual galaxies. 

\subsubsection{Inclination Errors}
Even when linewidths are measured very accurately,  
uncertainties are introduced in  correcting these 
velocity widths for inclination. Most of our optically-derived inclinations
should be good to $\pm$5$^{\circ}$ (MG), but for some galaxies, errors could
be larger due to the significant optical asymmetries, warps, 
or irregular outer isophotes which are
common in extreme late-type spirals. Such effects
may increase the scatter in 
TF by a few tenths of a magnitude (Franx \& de Zeeuw 1992).
However, we argue that there is unlikely to  be a systematic trend 
toward the overestimation
of rotational velocities from inclination effects. Optical inclinations 
are generally derived using the axial ratio of the observed isophotes, 
assuming that the intrinsic face-on galaxy isophotes would be circular. 
However, if a galaxy has intrinsically non-circular isophotes (as is 
often the case for very late-type spirals), even if 
it is observed face-on, the inclination 
of the galaxy will be measured to be non-zero; this effect should become 
less pronounced as the galaxy is turned edge-on. Therefore in most 
cases,  inclination errors should be in the sense of the 
inclinations being overestimated, and the 
galaxy's corrected rotational velocity being {\it underestimated}. 

In Fig.~4 we show a plot of $(\Delta TF)_{V}$ versus inclination.  
Here we see only a very slight correlation, which is small compared with the 
scatter, and we find TF deviators 
among the highest $i$ galaxies (i.e., the galaxies with the smallest
inclination corrections to their total linewidths). We note that one of 
our largest TF deviators (ESO~504-017) also has one of the smallest 
inclinations  ($i=30.8$) in our sample. 
However, we retain this galaxy in our analysis, since 
even for $i\approx60^{\circ}$ it would still fall more than 
2$\sigma$ below the standard $V$-band TF relation. 

\subsubsection{Distance Uncertainties} A major drawback of analyzing TF in a
nearby galaxy sample is that
deviations  from 
a simple linear Hubble flow may occur due to local peculiar 
velocities, leading to errors in distance determinations.  This 
is likely to be the largest source 
of uncertainty in our present analysis. Unfortunately, current
characterizations of the local peculiar velocity 
field are still incomplete and uncertain on small scales
(e.g., Shaya \etal\ 1992 and references therein). Nonetheless, we need 
to examine the possibility 
that some or all of the apparent offset of extreme late-type galaxies 
from the standard TF relation is due to errors in distances caused by
peculiar velocity motions.

One recent model of nearby  galaxy flows was
presented by Shaya \etal\ (1992). 
In a companion paper, Tully \etal\ (1992) used this model to
derive distances to 
many of the galaxy groups defined in the TNGC. As suggested by Tully 
\etal\ (1992), distances to other Local 
Supercluster galaxies may be determined by assigning the galaxy in 
question to a TNGC group. We therefore utilize this approach to 
estimate peculiar velocity-corrected distances for our combined 
ELT sample galaxies.

Using Tully \& Fisher (1987) and the TNGC, we were able to make 
group assignments to 39 of our combined ELT sample galaxies 
(Table~1). Tully \etal\ (1992) provide distances for 32 of these 
groups.
In addition, 5 of our galaxies lie within a 6$^{\circ}$ projected 
radius from the center of the Fornax cluster. For these objects, we adopt 
as the distance the current best-estimate for the distance to the 
Fornax cluster of 18.3~Mpc (Freedman \etal\ 1997; Silbermann \etal\ 1998). 
Bureau \etal\ (1996) have shown that normal galaxies 
within 6$^{\circ}$ of the Fornax center follow the same 
TF relation as other nearby clusters, so {\it a priori} we do not expect to see 
unusual deviations among these objects.

Adopting the group distances given in Table~1, we present a 
revised $V$-band TF relation 
in Fig.~5. Here we see that while there is some reduction in scatter,
there is still a mean offset of 0.69 magnitudes 
for the combined ELT sample.  Although the ELT sample plotted here
is smaller than in 
Fig.~1, we still find 9 galaxies that are underluminous by 
$>$2$\sigma$ compared with the prediction of the TF relation. 
Therefore, current peculiar velocity models cannot reconcile the 
apparent offset of  the bulk of extreme late-type spirals from the 
standard TF relation.  Given the uncertainty in our present knowledge 
of the local peculiar velocity field, we of course
cannot rule out that in a few individual  cases, distance errors may 
be responsible for the observed 
offsets. Overall, we estimate 
typical uncertainties in absolute magnitude due to distance errors 
of $\sim\pm$0.45 magnitudes, 
which is the mean difference in the absolute magnitudes determined by 
the linear Hubble flow method versus the group assignment method.

Another way of assessing the effect of distance errors on our TF 
comparison  is to examine the relationship between absolute magnitude and 
linear size ($A_{26}$) 
for our sample. Erroneous distances that cause errors in the 
derivation of absolute magnitudes will also cause errors in the derived 
linear sizes of the galaxies. In Fig.~6~$(a)$ we plot absolute $V$ 
magnitude for our combined ELT sample galaxies versus the logarithm of 
the linear diameter, $A_{26}$,
in kiloparsecs. Both of these quantities are derived assuming 
linear Hubble flow distances. In Fig.~6~$(b)$ we plot the same 
quantities, but this time we have assigned to each galaxy a revised 
distance assuming 
it adheres to the standard TF relation.  It is clear in 
Fig.~6~$(b)$ that forcing our galaxies onto the TF relation results in a 
very unrealistic distribution of linear sizes versus absolute 
magnitudes. 
Galaxies with $M_{V,i,b}\sim-18$ appear to span a factor of 12 is 
diameter, and the physically smallest galaxies in our sample would have 
to
span  a range of 
almost 6 magnitudes in brightness. Again this suggests that our 
observed offset from the TF relation cannot be fully accounted for by 
distance errors.

\subsection{Discussion}

We conclude that systematic errors are unlikely to be the source of the 
{\it global} offset of the combined ELT sample from the standard TF relation.
Summing in quadrature our expected sources of error, including distance 
uncertainties ($\pm$0.45 magnitudes), 
photometric
uncertainties ($\pm$0.33 magnitudes),
$W_{20}$ measurement uncertainties ($\pm$0.25 magnitudes), internal 
extinction uncertainties
($\pm$0.2 magnitudes),
and inclination effects  ($\pm$0.3 magnitudes), we estimate a typical
error budget for the comparison of any individual galaxy with TF 
of $\sim\pm$0.71 magnitudes in the $B$ band. 
In the $V$ band, we should also take into account 
possible
magnitude errors for the PT galaxies due to aperture corrections of
$\sim\pm$0.15 magnitudes (de Vaucouleurs \etal\ 1976), yielding a total 
uncertainty of $\pm$0.72 magnitudes for each individual galaxy. 
While we cannot rule out that in a few cases, distance 
errors, combined with observational uncertainties may produce spurious 
deviations from TF, the typical errors that we estimate are 
insufficient to account for the large number of galaxies we find that 
deviate by more than 2$\sigma$ from the standard TF relation. In addition,
we emphasize that errors in inclinations and internal
extinction estimates may have actually {\it decreased} any intrinsic offset 
of some of 
our combined ELT sample objects from TF. 

The type of observed offset from TF we see
among our extreme late-type spirals is
consistent with past indications of a curvature or offset
at the faint end of TF for other low surface brightness (LSB) and/or
low-luminosity galaxies in the sense of these galaxies being 
underluminous for a given value of $W_{20}$
(e.g., Romanishin \etal\ 1982; 
Carignan \& Beaulieu 1989; 
Persic \& Salucci 1991; Salpeter \& Hoffman 1996; Meurer \etal\ 
1996; Freeman 1997; Salucci \& Persic 1997; Walsh 1998).

MG have emphasized that there exists a significant range in the physical
properties of extreme late-type spirals (see also Paper~I). 
Consistent with this 
assertion, even if we assume that the majority of the combined ELT 
sample galaxies
are in accordance with the standard, linear TF relation
to within 
observational uncertainties,
this still leaves at least 9 galaxies in our combined ELT 
sample that lie more than 2$\sigma$
below the standard $V$-band TF calibration, even after corrections to 
distances for peculiar velocity motions (Fig.~5).
Therefore, {\it among the 
extreme late-type spirals there appear to exist examples of
rotationally-supported 
disk galaxies that are underluminous  for their observed rotational 
velocities}. 

Because very late-type galaxies can frequently have  dynamics
dominated more by the gaseous than  the stellar components 
(e.g., Jobin \& Carignan 1990; C\^ot\'e \etal\ 1991; Meurer \etal\ 
1996),
Milgrom \& Braun (1988) have suggested adding to the optical luminosity a
correction for the \mbox{H\,{\sc i}} ``luminosity'' or a ``baryonic
correction'' in order to
preserve the TF relation for certain galaxies. 

Using the PEGASE stellar evolutionary code of Fioc \& 
Rocca-Volmerange (1997), we estimate a typical
(${\cal M}/L$)$_{\star}$ ratio for extreme 
late-type spirals of $\sim$1 (see also Bica \etal\ 1988). Assuming
a helium fraction of Y=0.25, and negligible molecular gas contents, we 
apply to all of our combined ELT 
sample galaxies, as well as to the PT calibrator galaxies, a baryonic 
correction of the form:
\begin{displaymath}
L_{bary}=L_{V}+1.33{\cal M}_{\HIss}.
\end{displaymath}
\noindent We see in Fig.~7 that 
applying the baryonic correction can reconcile some but not most of the
combined ELT sample galaxies with the TF relation defined by the PT local 
calibrators.  We still see a global 
offset of the combined ELT sample of 0.63 magnitudes relative to the 
local calibrators, since the 
baryonic correction also 
shifts to brighter magnitudes the late-type
galaxies that anchor the faint-end of the local calibrator 
sample. In addition, the baryonic correction
does not significantly 
reduce the scatter in the TF relation for the combined ELT sample 
galaxies, and greatly increases the dispersion of the PT galaxies about 
the best-fit power law. 

In Fig.~8 and 9 we see that there exist  only weak
correlations between both \dark\ and mean disk surface brightness, 
$\bar\mu_{V}$, and the deviation from the TF relation,
$\Delta(TF)_{V}$ in the combined ELT sample.
Together these results suggest that while to some degree, our TF deviators 
appear to be galaxies that have been inefficient at converting their raw 
material into stars (cf. Hunter \& Gallagher 1986),  this cannot
fully account for 
the deviation from the TF
relation for most of our sample objects (see also Meurer \etal\ 1996).
These 
findings hint that the combined gas 
and stars in the inner galaxy 
is not what primarily determines $V_{max}$ of the rotation curves for
some of our extreme late-type spirals---i.e., these galaxies may be
dark matter-dominated even within the optical galaxy. Similar results 
have also emerged from recent rotation curve studies of these objects 
(e.g., Matthews \& Gallagher 1998). Such effects are
expected, for example, if the degree of luminous-to-dark
matter coupling decreases among low-mass galaxies (Salucci \& Persic 
1997).
We discuss this possibility further in the next section. 

\subsubsection{The Possible Role of Dark Matter}
Using the parameters given by Carignan \& Beaulieu (1989) and
Meurer \etal\ (1994,1996), we have overplotted on
Fig.~1~$(a)$ two additional small, faint disk galaxies:
NGC~2915 and 
DDO~154. It is clear that both of these galaxies deviate significantly 
from the standard $B$-band TF relation. Meurer \etal\ (1996) 
has discussed the 
likelihood
that the deviations of 
NGC~2915 and DDO~154 from TF are a consequence of the high dark
matter domination  of these galaxies. Could this also be the case 
for our TF deviators? 

In Fig.~10 we plot $\Delta(TF)_{V}$ versus $V$-band 
optical linear size ($A_{26}$) in kpc for our combined ELT sample. 
Here we see an apparent tendency
for the smallest galaxies to be the greatest 
deviators from the TF relation. Since for a gravitationally-supported disk, 
if all the mass is visible, then $V^{2}_{rot}r\propto L$, where 
$V_{rot}$ is the rotational velocity, $r$ is 
galactocentric distance, and $L$ is the luminosity. Therefore
smaller galaxies with a given $L$ should rotate faster. The observed
adherence of most disk galaxies to a universal TF relation 
implies that a given amount of 
optically
luminous matter is always radially scaled to produce a set, corresponding 
rotational velocity---i.e. that the product of the galaxy surface density 
$\bar\Sigma$ and the mass-to-light ratio 
${\cal M}/L$ is a constant (e.g., Zwaan \etal\ 1995).
However, this trend may break down among 
some of the smallest spirals (e.g., Kormendy 1990; 
Salucci \& Persic 1997).

Our present findings suggest that  at a given rotational 
velocity, the smallest galaxies are physically distinct from the largest 
ones, and that total matter distributions (or $\bar\Sigma$) are not always 
simply radially scaled as to preserve a linear TF relation.
A natural interpretation of this is that the most severely TF-deviating 
galaxies in our sample are likely to be galaxies where the luminous 
matter distribution (\HI\ + stars)
contributes the least to establishing 
the total rotational velocity of the galaxy, and 
where the luminous matter distributions are compact relative to 
the extent of the  dark matter halo. Such phenomena are predicted by 
dynamical models for low-luminosity disks (Levine \& Sparke 1998), 
galaxy scaling laws (Kormendy 1990; see also Freeman 1997),
and for the
models of dark and luminous matter coupling in galaxies proposed by
Salucci \& Persic (1997). Salucci \& Persic (1997) have also shown that 
their model predicts a 
downturn in the TF relation among low-luminosity, moderate-to-low 
surface brightness galaxies that is 
consistent with that observed in
the extreme late-type spiral sample of Matthews \etal\ (1997). 

Further indirect evidence that dark matter may play a role in the deviation of 
some of our sample galaxies from the standard  TF relation comes from 
Fig.~11, where we plot absolute $V$ magnitude for our combined ELT sample 
galaxies versus $\Delta(TF)_{V}$. Here we see some indication 
that deviation from TF 
appears to increase with decreasing luminosity in our sample. This 
hints that the trend we are observing is indicative of a {\it 
curvature} rather than a global offset at the faint end of the TF 
relation (see also Salucci \& Persic 1997). This type of offset is 
consistent with the predictions of 
galaxy scaling laws, which show that the density of 
dark matter halos increases with decreasing galaxy luminosity, 
resulting in increasingly dark matter-dominated inner disks in fainter 
spirals (Kormendy 1990). Previous results of aperture synthesis studies
have indicated that the 
increasing importance of dark matter starts to become evident at 
$M_{V}\approx$-17.5 (Skillman \etal\ 1987; Lake \etal\ 1990; Meurer 
\etal\ 1996), approximately the same
point at which we see our sample 
galaxies beginning to fall consistently below the TF relation. 

The true form of the TF relationship has important implications for 
testing 
the robustness of galaxy formation models.  For example, Dalcanton 
\etal\ (1997) predict only a slight curvature in the
TF relationship for their models of
disk galaxies. They, however, find good agreement with observations
in that they see only a very weak dependence of the TF relationship on
disk surface brightness (cf., Fig.~9).  Recent semi-analytic models by
Mo \etal\ (1998), in which the authors seek to present 
a consistent picture of galaxy 
formation within a given cosmology, also show only a very slight curvature
down from a linear TF relationship at low masses. However,
a wider range of downward curvatures is seen, for example,
in some of the models  of  Somerville \& Primack (1998). While
detailed comparisons between global characteristics of low mass
disk galaxies and predictions of formation models are beyond the
scope of this paper, this issue clearly warrants future attention.

\subsubsection{Comparison with Other Recent Observational Studies}
The results we present here for the TF relation of extreme late-type 
spirals may seem at first
glance contrary to the conclusions reached in two recent investigations
of the TF relation for LSB spirals: Sprayberry {\it et al.} (1995) and Zwaan
{\it et al.} (1995).  These authors find that their LSB spiral samples
follow the {\it same} TF relation as high surface brightness
spirals (see also de Blok 
\etal\ 1996).
We argue that the difference is due primarily 
to the examination of a somewhat 
different class of galaxies than the present study:
the predominantly ``large'' LSB galaxies
versus the extreme late-type spirals. 

The LSB galaxy samples studied by Zwaan \etal\ (1995), Sprayberry \etal\ 
(1995) and de~Blok 
\etal\ (1996) are mainly composed of larger scale-length spirals with 
typical luminosities and \HI\ masses larger than in our 
combined ELT sample. For example, in the de Blok \etal\ (1996) sample, 
the mean log(\mhi) is 9.23$\pm$0.27~\msun\ and the mean absolute $B$ 
magnitude is $\bar M_{B}$=-17.53$\pm$0.88 (for
$H_{0}$=75~\kms~Mpc$^{-1}$), compared with a  mean 
log(\mhi) of 
8.78$\pm$0.50~\msun\ and $\bar M_{B}$=-16.56$\pm$1.14
in the combined ELT
sample.  
In the Sprayberry \etal\ (1995)
sample, $\bar W_{20}$=286$\pm$84~\kms\ (before inclination correction) 
while $\bar W_{20}$/${\rm sin}~i$=146$\pm$52~\kms\ 
for 
our combined ELT sample. Because many of our combined ELT sample 
galaxies lie outside of the $M_{B}$, $W_{20}$,
\mhi, and $A_{26}$ domains that typify the 
samples of 
Sprayberry {\it et al.} (1995), Zwaan
{\it et al.} (1995), and de Blok \etal\ (1996), and since in many cases
we find the most significant departures from TF for the smallest and 
faintest disks 
(see also Salpeter \& Hoffman 1996),
this may explain why 
these workers did not find similar evidence for deviations from the TF 
relation in their samples. 

\section{Summary and Conclusions}
Using new and previously published \HI\ data, together with the 
photometry of Matthews \& Gallagher (1997; MG), we have explored
the $B$ and $V$-band Tully-Fisher (TF) relation for 44
extreme late-type 
spiral galaxies. Relative to the TF relation defined by 
the local calibrator sample of 
Pierce \& Tully (1992),  we find offsets of 0.64 and 0.82
magnitudes in $B$ and $V$, respectively,
for our sample of extreme late-type
spirals. For 37 of our sample galaxies, following Tully \etal\ (1992),
we have derived revised 
distances based on assignments to galaxy groups whose distances have 
been previously determined using models for local peculiar velocity 
flows. The extreme late-type spirals 
with corrected distances still show a mean offset from TF of 
0.69 magnitudes in $V$. 
Thus within the limits of present models,
local peculiar motions cannot account for the 
effects we observe. 
Adopting distances to our sample galaxies by assuming they do 
lie on TF results in an unrealistic distribution of absolute magnitude 
 versus linear diameter. This is further evidence that distance errors 
are not the source of the observed departure of many extreme late-type
spirals from TF. 
Therefore, our data show evidence for the existence 
of populations of rotationally
supported disk galaxies which lie below the
normal, linear TF relation---i.e., these galaxies appear to be
optically under-luminous for their rotation speeds.

While offsets of individual galaxies from the TF relation 
of up to $\pm$0.72 magnitudes 
may be explicable as artifacts of
observational uncertainties, or individual distance determination 
errors, these effects cannot account for 
the observed global offset of our sample.
Even after correcting distances for local peculiar motions, 
we still find 9 extreme late-type spirals
that deviate by more than 2$\sigma$ (i.e. $>$1.16 magnitudes) from the 
$V$-band 
TF relation defined by the local calibrator sample.

We find that deviation from the TF relation, $\Delta(TF)$, 
appears to increase
with
decreasing linear optical size and decreasing luminosity
in our sample, which 
suggests that the greatest TF deviators are galaxies
where the luminous matter contents are small and compact relative
to the size and extent of their dark matter halos and that there may be 
curvature at the faint end of the TF relation. We see only very weak
correlations between  $\Delta(TF)$ and inclination or
mean disk surface brightness, and no obvious correlation between 
$\Delta(TF)$ and
$B - V$ color.
Including the \HI\ flux as 
part of the total luminosity cannot reconcile most of these 
galaxies with the TF relation, and $\Delta(TF)$ does not
correlate strongly
with increasing \dark. These trends suggest 
that the TF deviators are not simply 
galaxies that have retained large gas reservoirs due to slow evolution,
and hints that the TF deviators may be galaxies where the sum 
of the visible gas 
and stars contributes little to the determination of $V_{max}$ of the 
rotation curve---i.e. 
they may be dark matter dominated even within the optical galaxy. 
Observations of these low mass 
disk galaxies may therefore
provide useful constraints on galaxy formation models.

\acknowledgements{This work made use of data obtained at the \nan\ 
Radio Observatory. The \nan\ Observatory is the department {\it 
Unit\'e
Scientifique \nan} of the {\it Observatoire de Paris} and is associated 
with
the French {\it Centre National de Recherche  Scientifique} (CNRS) as
the {\it Unit\'e de Service et de Recherche} (USR), No. B704. The
Observatory also gratefully acknowledges the financial support of the
{\it R\'egion Centre} in France. We acknowledge useful 
discussions with M. J. Pierce and A. Yahil
on  aspects of this work and thank D. M. Peterson
for a critical reading of an earlier version of this manuscript.
JSG's work in this area benefitted from his participation
in a workshop at the Aspen Center for Physics.
LDM is supported by a graduate internship with the Wide
Field and Planetary Camera 2 Investigation Definition Team, which is
supported at the Jet Propulsion Laboratory (JPL) via the National
Aeronautics and Space Administration (NASA) under contract No.
NAS7-1260. This research made use of the NASA/IPAC
Extragalactic Database (NED), operated by JPL under contract with NASA.}

\clearpage

\pagestyle{empty}
\footnotesize
\begin{deluxetable}{llc}
\tablewidth{15pc}
\tablecaption{}
\tablehead{
\colhead{Galaxy}          & \colhead{Group} & \colhead{D(Mpc)} }

\startdata
{\bf Paper~I Galaxies} & \phm{11-0} & \phm{11} \nl
\tablevspace{1mm}
ESO~547-020&	51-0+4 & 22\nl
ESO~418-008 & Fornax & 18.3\nl
ESO~418-009 & Fornax & 18.3\nl
ESO~482-005 & 51-4 & 20 \nl
ESO~358-015 & Fornax & 18.3\nl
ESO~358-020 & Fornax & 18.3\nl
ESO~548-050 & 51-4 & 20 \nl
ESO~549-002 & 53-0$^{*}$ &  \nodata \nl
ESO~358-060 & Fornax & 18.3 \nl
ESO~359-016 & 51-0+18 &18  \nl
ESO~359-029 & 53-13 &  17  \nl
ESO~359-031 & 53-13 & 17 \nl
ESO~422-005 & 51-10 & 18 \nl
ESO~305-009 & 53-10 & 12 \nl
ESO~487-019 & 34-2 & 20 \nl
ESO~488-049 & 34+1$^{*}$ & \nodata \nl
ESO~425-008 & 34-1 & 21  \nl
AM0605-341 & 53-14 & 9 \nl
ESO~431-015 & \nodata &\nodata  \nl
ESO~497-007 & 31-0+12 & 30 \nl
ESO~500-032 & 31-0+2 & 30 \nl
ESO~502-016 & 22-0$^{*}$ & \nodata \nl
ESO~438-005 & 22-6 & 14 \nl
ESO~504-017 & 22-4 & 17 \nl
ESO~505-013 & 22-1 & 17 \nl
ESO~380-025 & 23-5 & 24 \nl
ESO~443-079 & 11-22 & 16 \nl
ESO~443-080 & 11-22 & 16 \nl
ESO~444-033 & 11-32 &20 \nl
ESO~446-053 & 16-0 & 10 \nl                                                  
\tablevspace{1mm}
{\bf MG Galaxies} & \phm{11-1} & \phm{11} \nl
\tablevspace{1mm}
SCG0448-39 & \nodata & \nodata \nl
ESO~552-066 & 34-3 & 22 \nl
NGC~2131 & 34+1$^{*}$ & \nodata \nl
ESO~377-017 & 31-10 & 27 \nl
ESO~503-022 & 22+4$\*$  &\nodata  \nl
ESO~504-010 & 22-14 & 17 \nl
ESO~440-039 & 20-5+4 & 21 \nl
ESO~440-049 & 22-5+4 & 21 \nl
ESO~508-034 & 11-22 & 16 \nl
ESO~445-007 & 11-0+22 & 15 \nl
ESO~510-026 & 11-0 & 28.5 \nl
ESO~318-024 & 54+4 & 9 \nl
ESO~440-004 & 22-4 & 17 \nl
ESO~504-025 & 22-4 & 17 \nl
ESO~507-065 & 11-23+22$^{*}$ &\nodata  \nl
ESO~443-083 & 11+32$^{*}$ & \nodata \nl
  \nl
\enddata

\tablenotetext{*}{Denotes group for which Tully {\it et al.} (1992) do not 
provide a distance estimate.}

\tablecomments{The first column is galaxy name from 
 the combined 
ELT sample; the second column give corresponding assignment of the galaxy
to the galaxy groups defined by 
Tully (1988). Galaxies within a 6$^{\circ}$ projected radius around the 
center of the Fornax cluster were assumed to be members of Fornax 
(Bureau {\it et al.} 1996). The last column gives the revised 
distance assigned to the galaxy based on the group assignment. These 
distances are used in Fig.~5.}
\end{deluxetable}

%

\begin{figure}
\figurenum{1a}
\epsscale{0.7}
\plotone{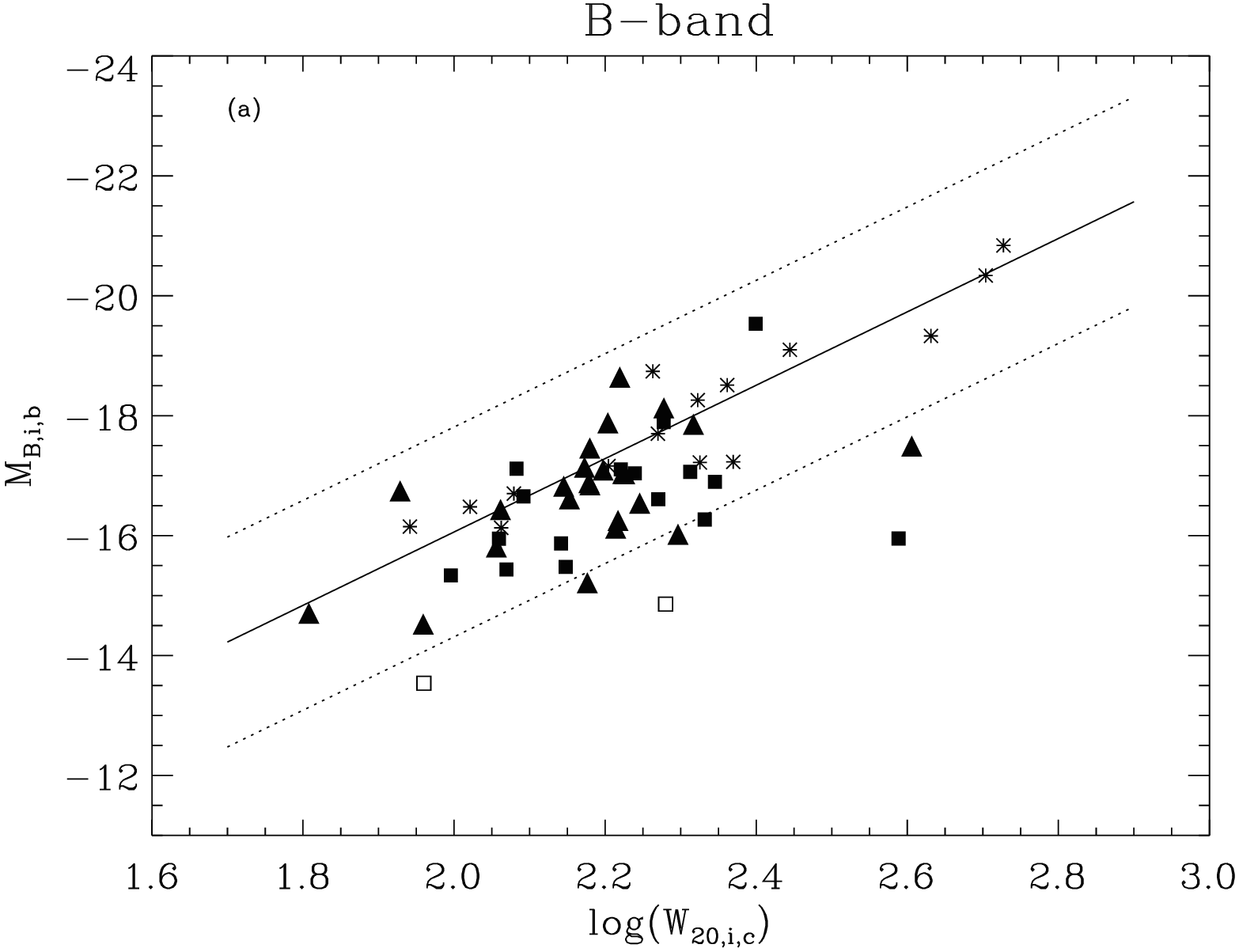}
\caption{$B$-band Tully-Fisher relation for the combined 
ELT sample and the local calibrator sample of PT.
Triangles are galaxies from Paper~I; filled squares are 
17 additional extreme late-type spirals from 
MG; asterisks are the PT sample. Distances to the combined ELT sample 
galaxies are derived using a simple linear Hubble flow with 
$H_{o}$=85~\kms~Mpc$^{-1}$. The axes are the 
logarithm of the observed \HI\ linewidth at 20\% peak maximum, in \kms, 
corrected for turbulence and inclination, versus the $B$-band 
absolute magnitudes, corrected for Galactic and internal extinction, as 
described in the Text.
In Panel~$(a)$, the open 
squares represent the galaxies DDO~154 (on the left) and NGC~2915 (on 
the right; see Sect.~2.5.1). The solid 
line represents a least squares fit to the PT data, and the dotted 
lines indicate  the 2$\sigma$ scatter.}
\end{figure}

\begin{figure}
\epsscale{0.7}
\figurenum{1b}
\plotone{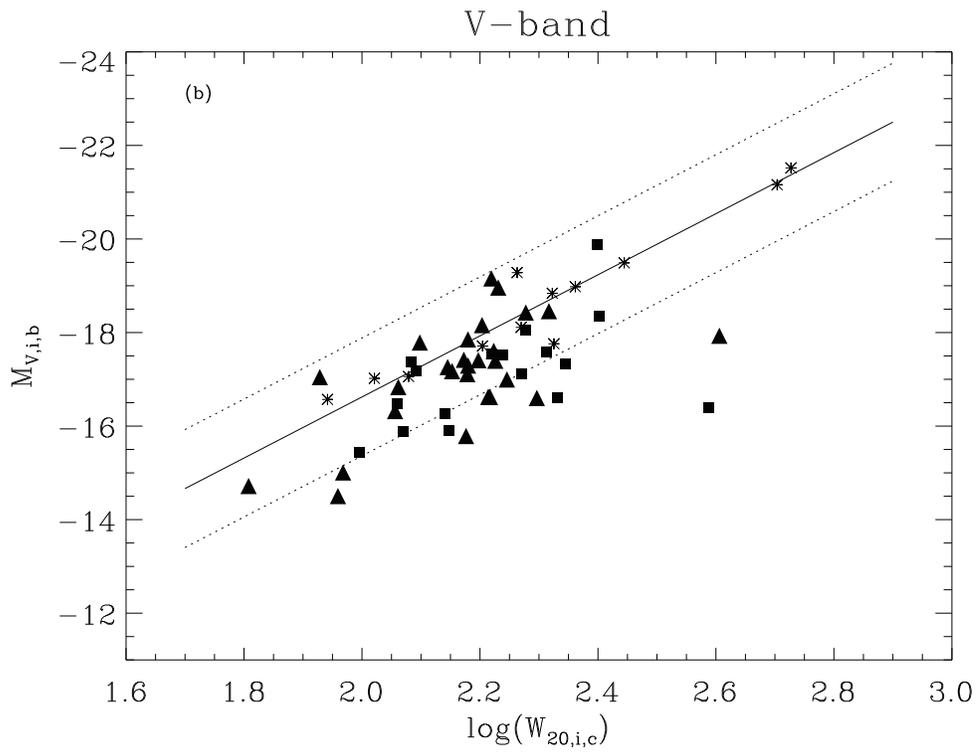}
\figcaption{$V$-band Tully-Fisher relations for the combined 
ELT sample and the local calibrator sample of PT. See Figure 1a 
for details.}
\end{figure}

\begin{figure}
\figurenum{2}
\epsscale{0.7}
\plotone{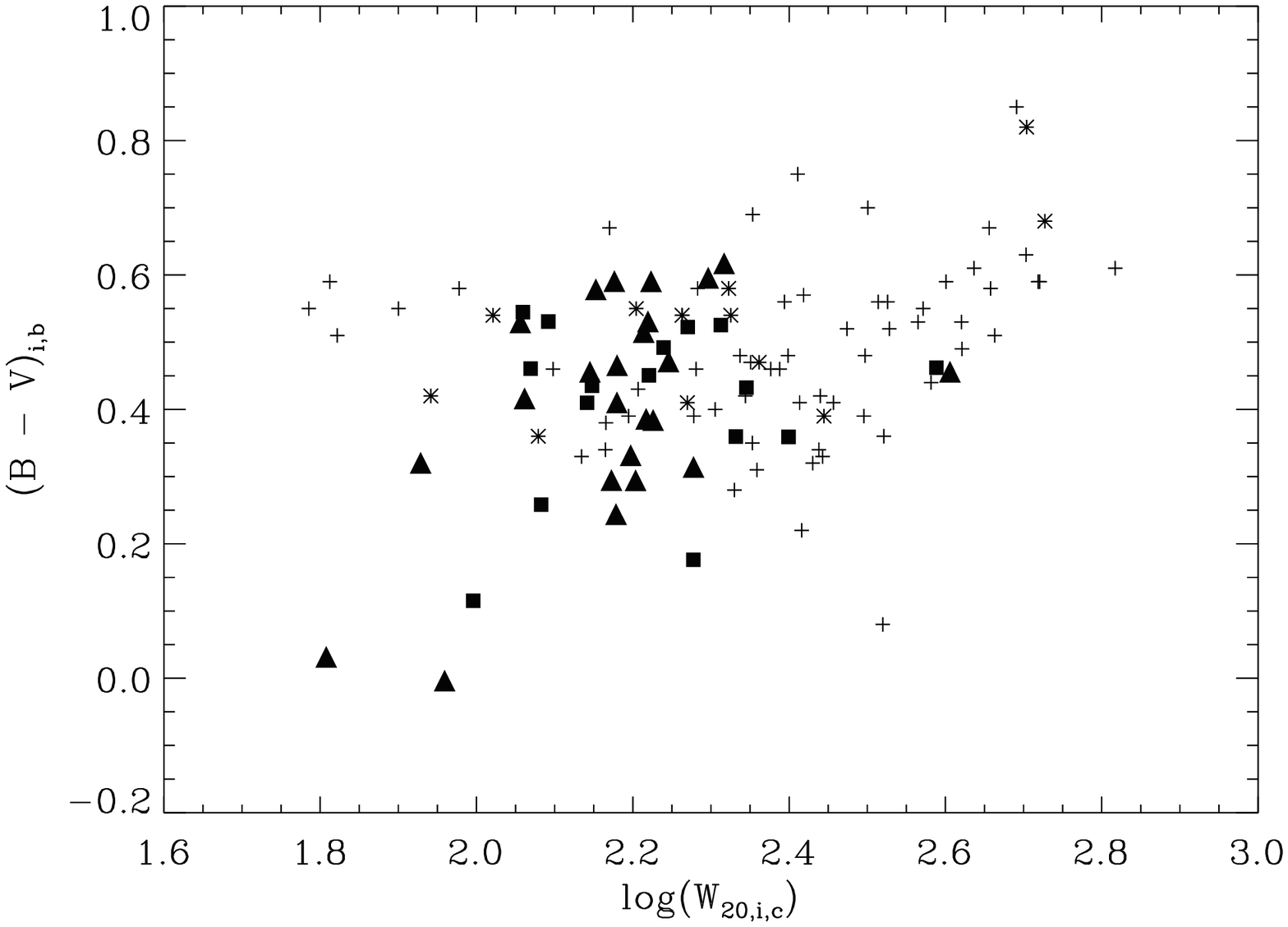}
\figcaption{The logarithm of the observed \HI\ linewidth at 20\% peak 
maximum, in \kms,
corrected for turbulence and inclination, versus $B - V$ color, 
corrected for internal and Galactic extinction. Pluses are 
galaxies from the TNGC; the remaining symbols are as in Fig.~1.}
\end{figure}

\ 
\begin{figure}
\figurenum{3}
\epsscale{0.7}
\plotone{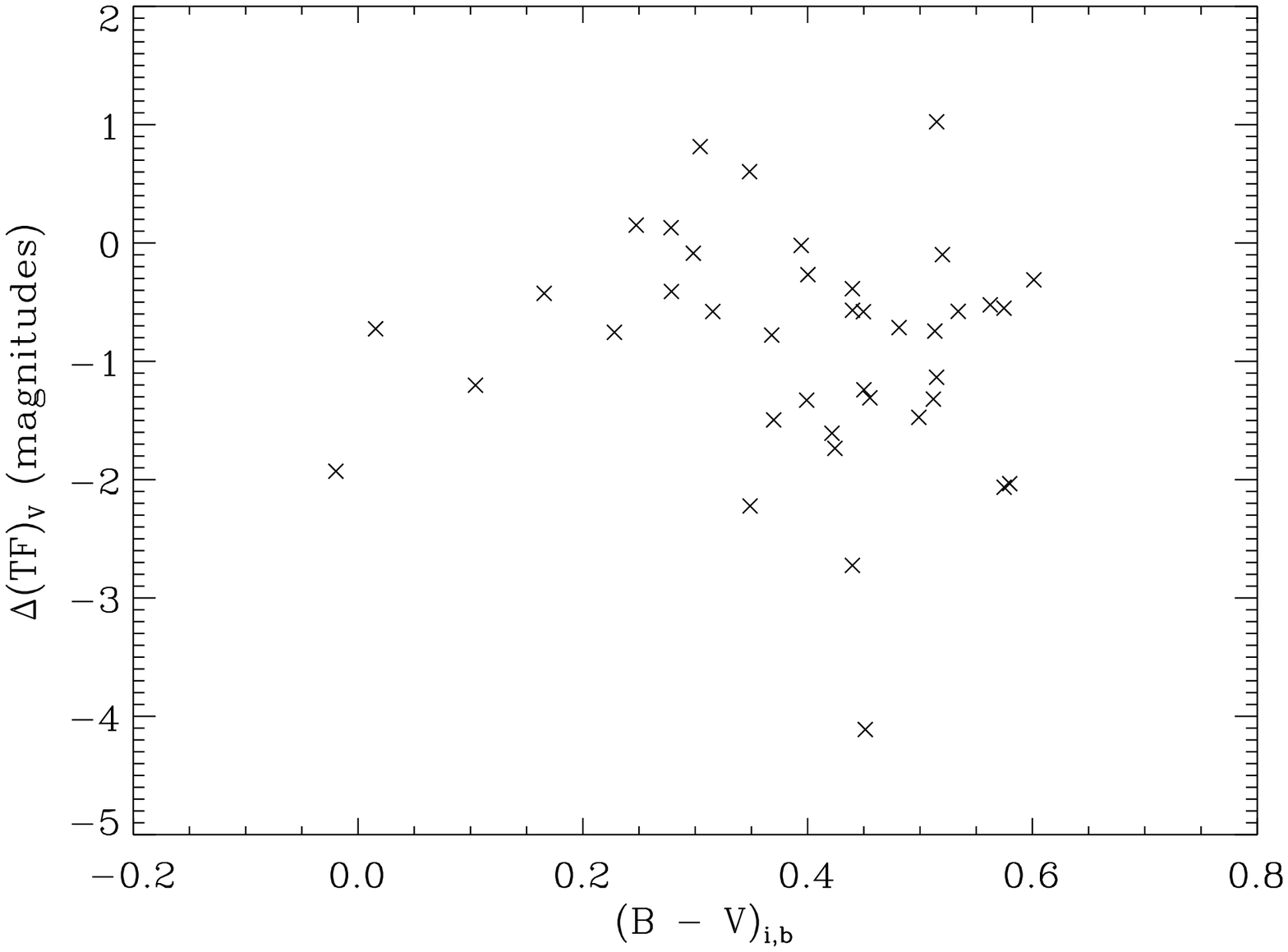}
\figcaption{Deviation of the combined ELT sample
from the standard $V$-band Tully-Fisher relation, in magnitudes,
versus $B - V$ color, corrected for Galactic and internal extinction.}
\end{figure}

\begin{figure}
\figurenum{4}
\epsscale{0.7}
\plotone{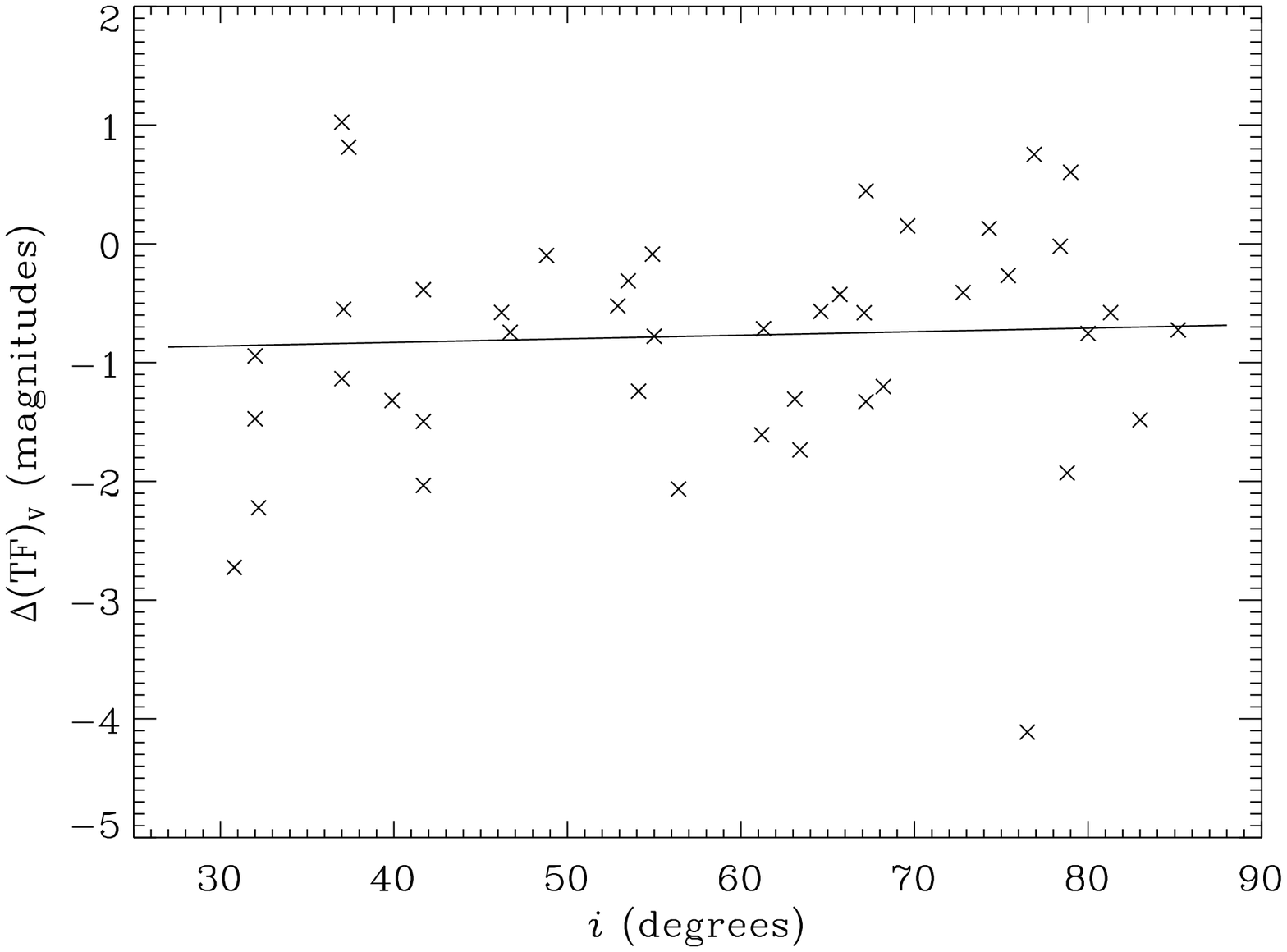}
\figcaption{Deviation of the combined ELT sample 
from the standard $V$-band Tully-Fisher relation, in magnitudes, 
versus inclination, in degrees. The solid line shows a least squares 
fit to the data.}
\end{figure}

\begin{figure}
\figurenum{5}
\epsscale{0.7}
\plotone{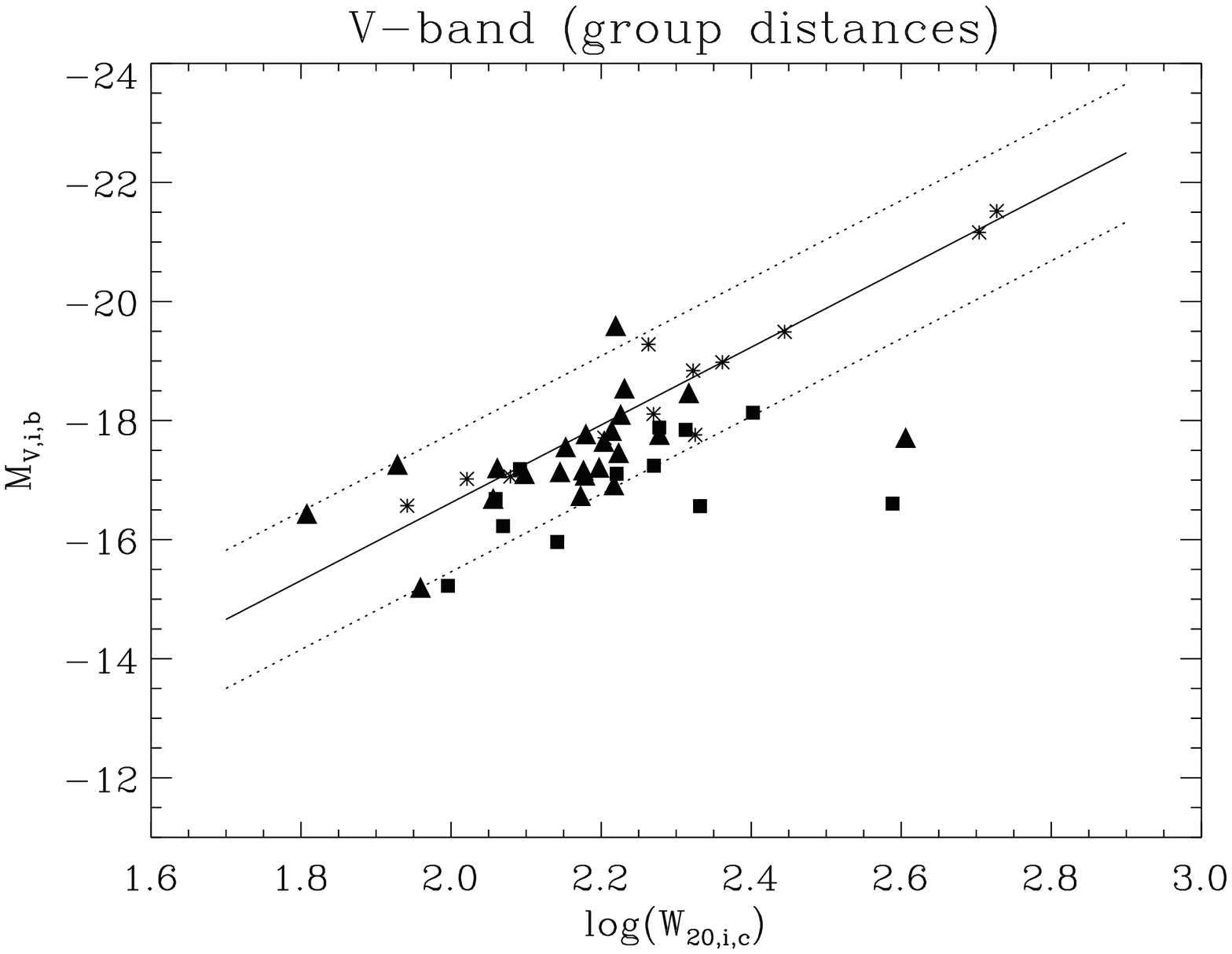}
\figcaption{Same as Fig.~1~$(a)$, except that distances to the combined ELT 
sample galaxies have been derived using galaxy group assignments (see 
Sect.~2.4.5).}
\end{figure}

\begin{figure}
\figurenum{6a}
\epsscale{0.7}
\plotone{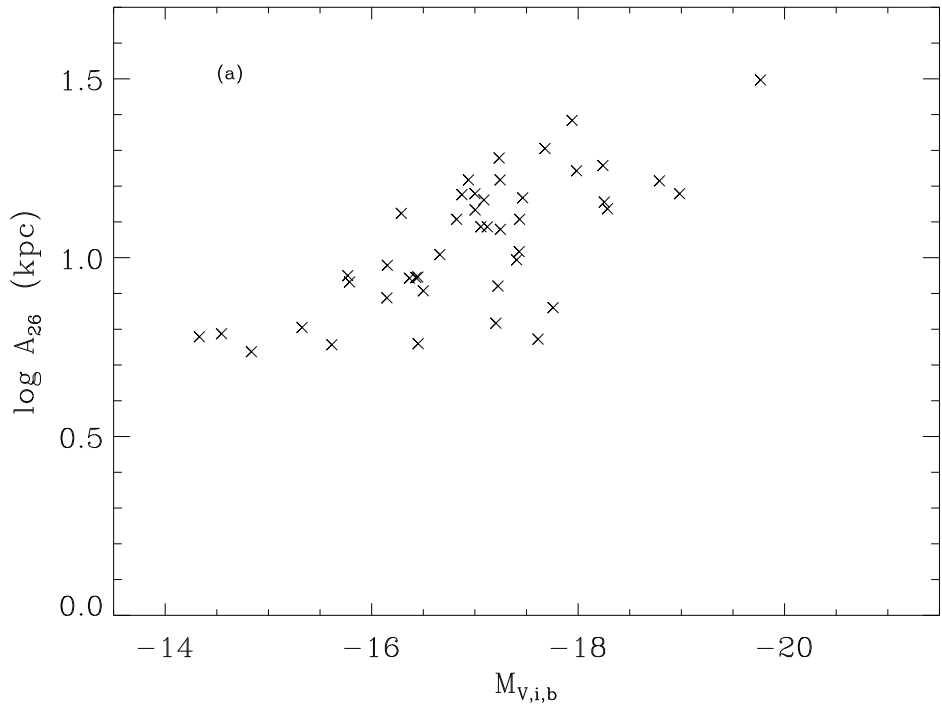}
\figcaption{The relationship between absolute $V$ 
magnitude, corrected for internal and external extinction, and the 
logarithm of the optical 
linear size (in kiloparsecs)
at the 26 magnitude per arcsecond
squared isophote for the combined ELT sample galaxies. Panel~$(a)$ 
assumes distances derived from a linear Hubble law.}
\end{figure}

\begin{figure}
\figurenum{6b}
\epsscale{0.7}
\plotone{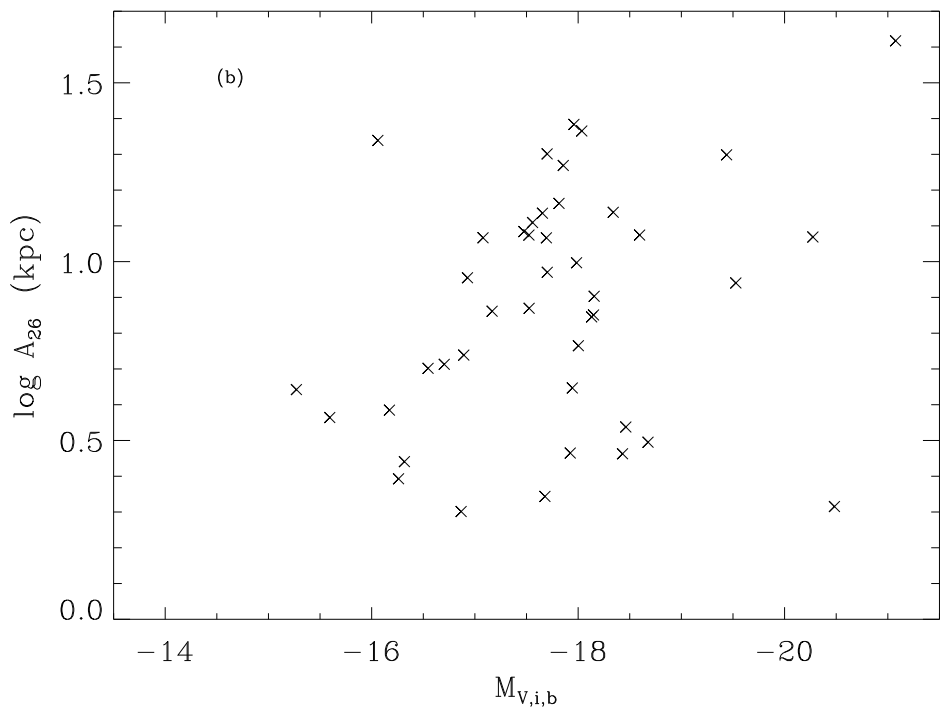}
\figcaption{Same as Figure 6a but panel~$(b)$ shows 
the resulting relation if distances to the galaxies are derived 
by assuming the galaxies adhere to the standard $V$-band TF relation 
shown in Fig.~1~$(b)$.}
\end{figure}

\begin{figure}
\figurenum{7}
\epsscale{0.7}
\plotone{ 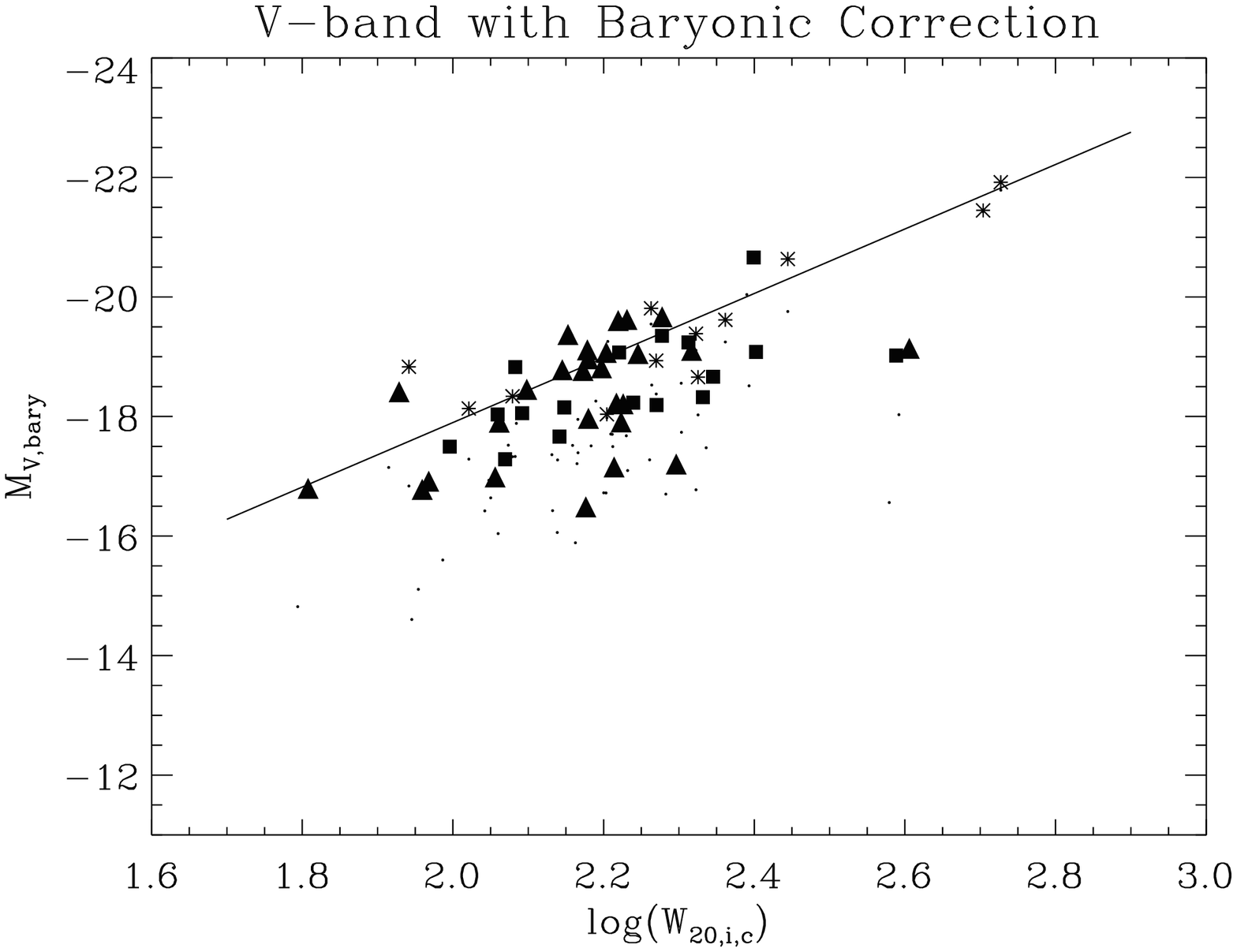}
\figcaption{Same as Fig.~1~$(b)$, except that baryonic corrections (see 
Sect.~2.5) have been applied to the luminosities of both the PT and the 
combined ELT sample galaxies. Small dots indicate 
positions of galaxies before the baryonic correction. The solid line 
is a least squares fit to the PT data. Distances to the combined ELT 
sample galaxies are derived assuming a linear Hubble law.}
\end{figure}

\begin{figure}
\figurenum{8}
\epsscale{0.7}
\plotone{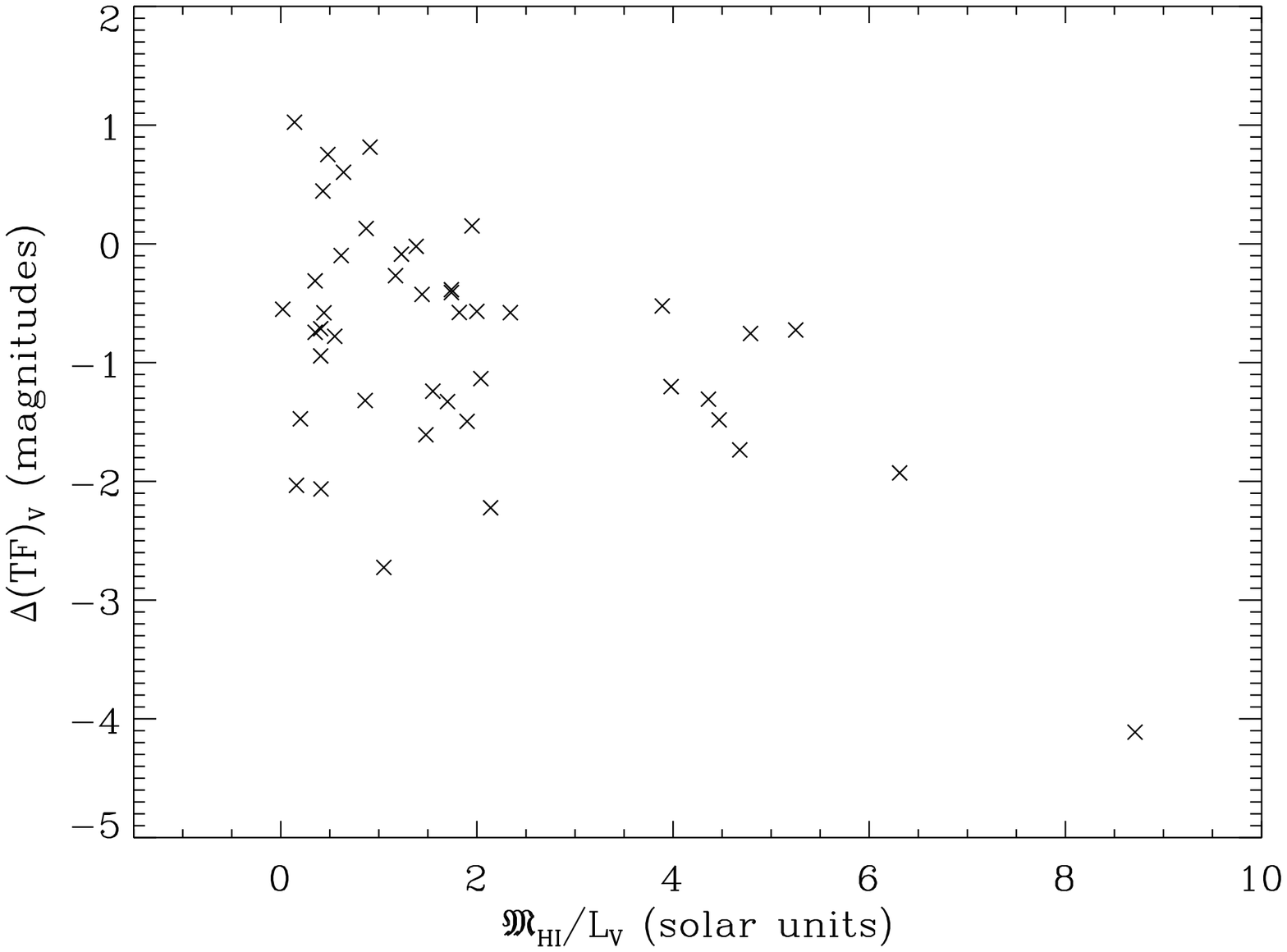}
\figcaption{Deviation of the combined ELT sample 
from the standard $V$-band Tully-Fisher relation, in magnitudes 
(assuming distances derived from a linear Hubble flow),
versus the ratio of the neutral hydrogen 
mass to the optical $V$-band luminosity, in solar units.}
\end{figure}

\begin{figure}
\figurenum{9}
\epsscale{0.7}
\plotone{ 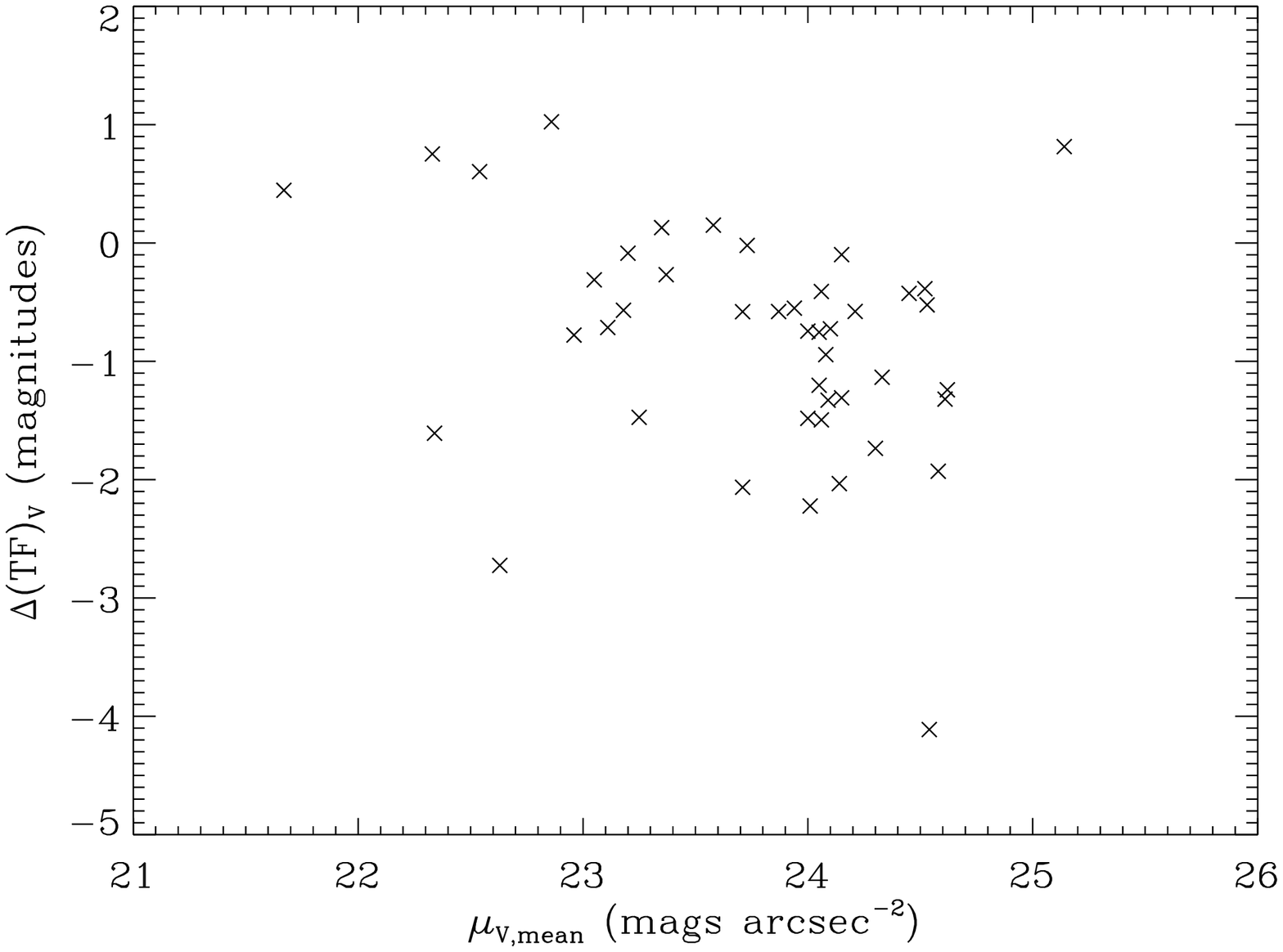}
\figcaption{Deviation of the combined ELT sample
from the standard $V$-band Tully-Fisher relation, in magnitudes 
(assuming distances derived from a linear Hubble flow),
versus the mean, observed $V$-band disk surface brightness,
in magnitudes per arcsecond squared. Mean disk surface brightnesses are 
taken from MG.}
\end{figure}

\begin{figure}
\figurenum{10}
\epsscale{0.7}
\plotone{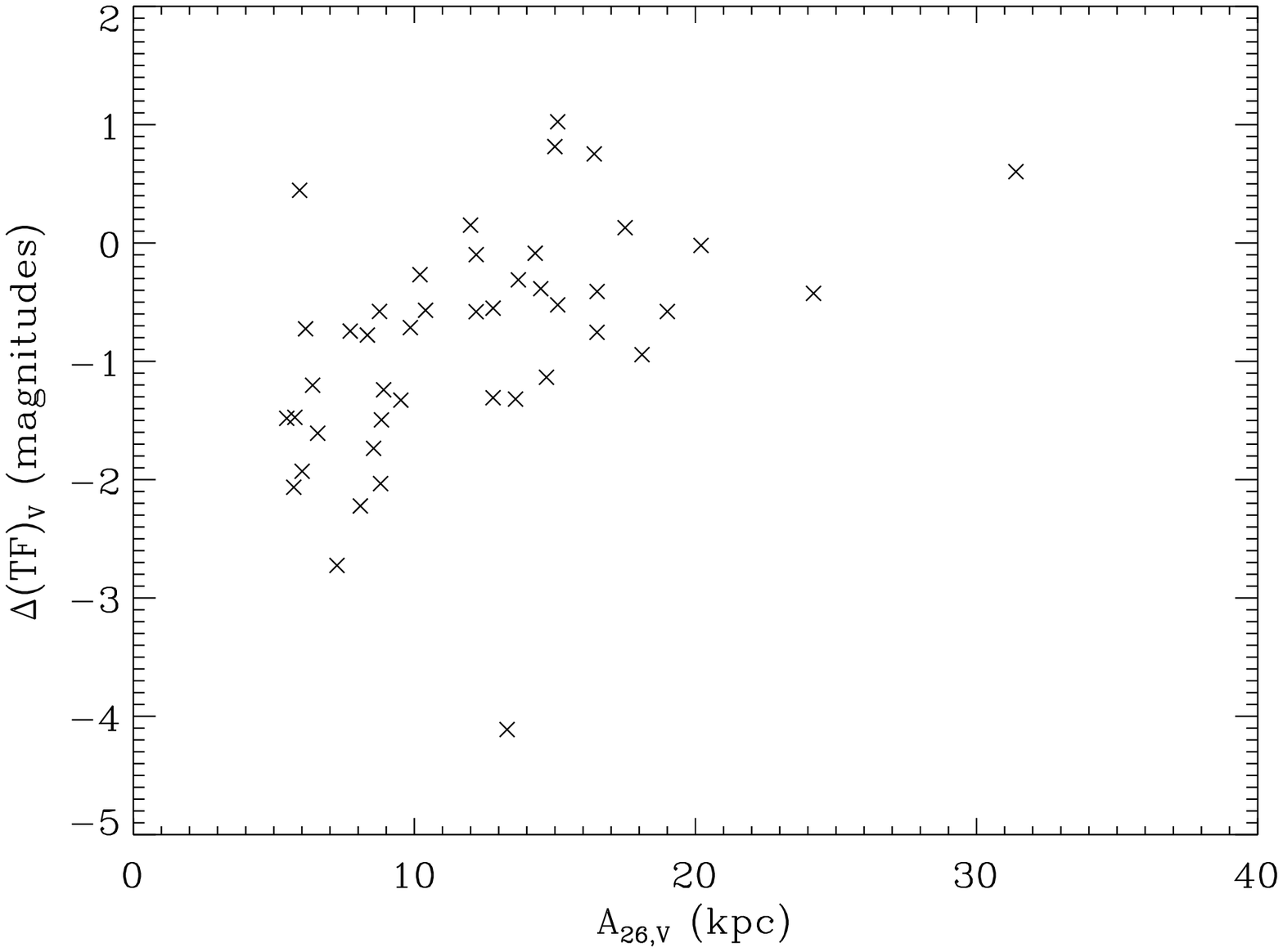}
\figcaption{Deviation of the combined ELT sample 
from the standard $V$-band Tully-Fisher relation, in magnitudes 
(assuming distances derived from a linear Hubble flow),
versus the $V$-band optical linear diameter (in kiloparsecs) 
at the 26 magnitude per arcsecond
squared isophote. Distances using a linear Hubble law are assumed for 
the combined ELT sample galaxies. The optical linear sizes are derived 
using the angular diameters of MG.}
\end{figure}

\begin{figure}
\figurenum{11}
\plotone{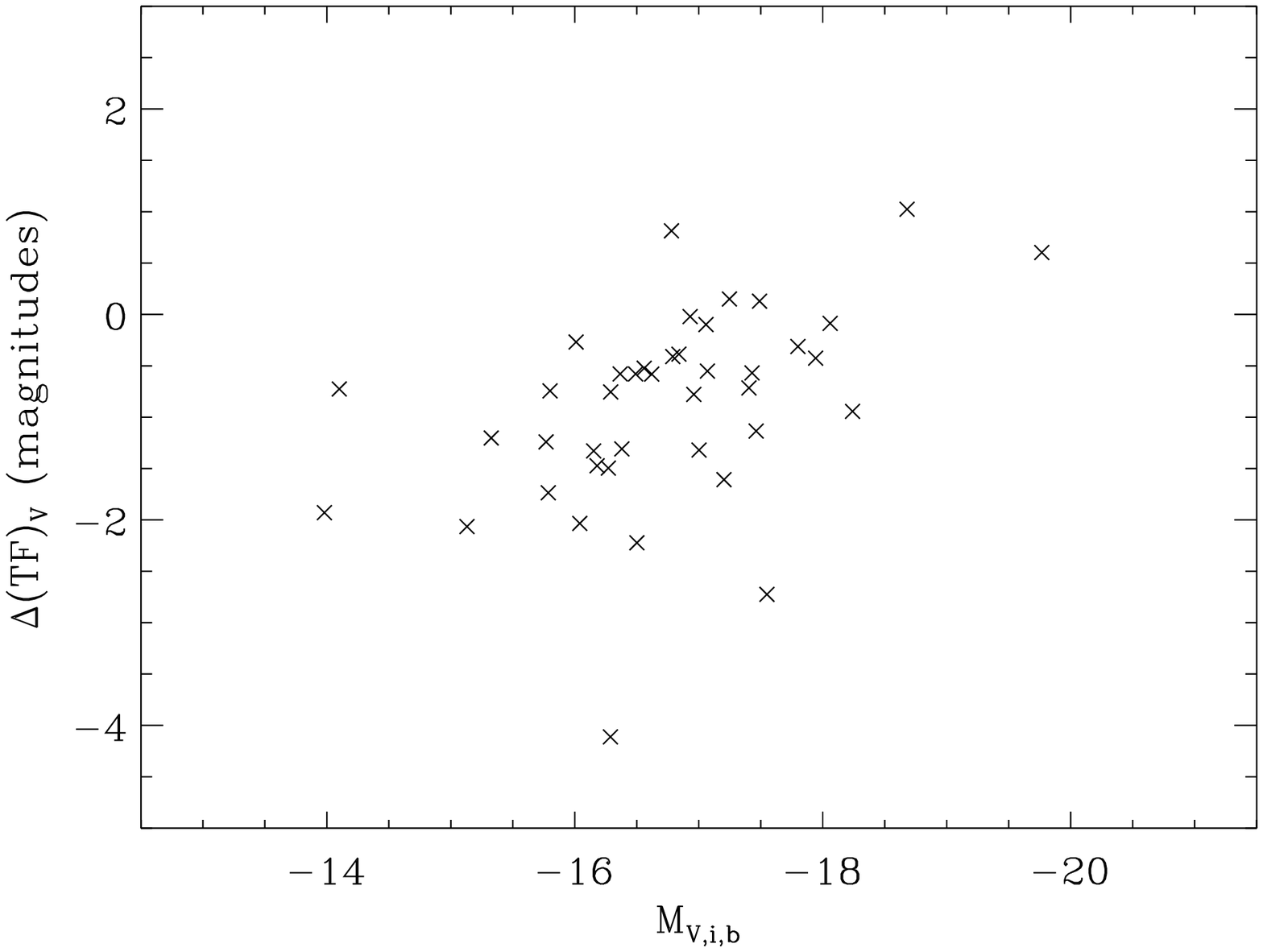}
\epsscale{0.7}
\figcaption{Deviation of the combined ELT sample
from the standard $V$-band Tully-Fisher relation, in magnitudes 
(assuming distances derived from a linear Hubble flow),
versus the absolute $V$-band magnitude, corrected for internal and 
external extinction. }
\end{figure}

\end{document}